\documentclass[graybox]{svmult}

\usepackage{mathptmx}       % selects Times Roman as basic font
\usepackage{helvet}         % selects Helvetica as sans-serif font
\usepackage{courier}        % selects Courier as typewriter font
\usepackage{type1cm}        % activate if the above 3 fonts are
\usepackage{makeidx}         % allows index generation
\usepackage{graphicx}        % standard LaTeX graphics tool
\usepackage{multicol}        % used for the two-column index
\usepackage[bottom]{footmisc}% places footnotes at page bottom
\usepackage{symb}

\makeindex             % used for the subject index
\begin{document}

\title*{BEAMS: separating the wheat from the chaff in supernova analysis}
\titlerunning{BEAMS}
\author{Martin Kunz, Ren\'ee Hlozek, Bruce A. Bassett, Mathew Smith, James Newling and Melvin Varughese}
%\authorrunning{M. Kunz, R. Hlozek, B.A. Bassett {\em et al.}}
\authorrunning{M. Kunz, R. Hlozek, B.A. Bassett, M. Smith, J. Newling and M. Varughese}
\institute{Martin Kunz \at D\'epartement de Physique Th\'eorique and Center for Astroparticle Physics, Universit\'e de Gen\`eve, Quai E.\ Ansermet 24, CH-1211 Gen\`eve 4, Switzerland, \email{Martin.Kunz@unige.ch}}
%\and Name of Second Author \at Name, Address of Institute \email{name@email.address}}
\maketitle

\abstract*{We introduce Bayesian Estimation Applied to Multiple Species (BEAMS), an algorithm designed
to deal with parameter estimation when using contaminated data. We present the algorithm and demonstrate
how it works with the help of a Gaussian simulation. We then apply it to supernova data from the
Sloan Digital Sky Survey (SDSS), showing how the resulting confidence contours of the cosmological parameters shrink 
significantly.}

\abstract{We introduce Bayesian Estimation Applied to Multiple Species (BEAMS), an algorithm designed
to deal with parameter estimation when using contaminated data. We present the algorithm and demonstrate
how it works with the help of a Gaussian simulation. We then apply it to supernova data from the
Sloan Digital Sky Survey (SDSS), showing how the resulting confidence contours of the cosmological parameters shrink 
significantly.}

\section{Introduction}
\label{sec:intro}

As demonstrated by the 2011 Nobel prize in physics, supernovae of type Ia (SN-Ia) are one of the most important
tools to study the expansion history of the universe \cite{riess_acceleration,Perlmutter_acceleration}. 
Supernovae are exploding stars that can be seen at extremely
large distances. The most distant supernova currently known (designated SN 19941, a type IIn supernova) is at a 
distance of 11 billion light years \cite{Cooke:2009ds}.
Due to the finite speed of light, a signal from a very distant source takes a very long time reach us, in this case
some 11 billion years (the universe itself is about 13.7 billion years old \cite{Komatsu:2010fb}).
During the time that the light of the explosion travels towards us, the universe expands and the wavelength of the light
is redshifted. By looking at spectral lines, it is possible to measure by how much the frequency changed. In the case of 
SN 19941the redshift is $z = 2.36$, i.e. the wavelength of the light was stretched by a factor of 2.36. This also implies that the
universe has grown by a factor of $(1+z)=3.36$ over the last 11 billion years.

In this way it is possible to map the expansion history of the universe and to compare it to predictions from 
General Relativity (GR). But in GR, space-time is curved, which makes the definition of distances somewhat
tricky. One way to do this assumes that we know the intrinsic luminosity of an object. Such an object is called
a {\em standard candle}, and astrophysical research has shown that SN-Ia are fairly good standard candles
after some data processing. If we have a standard candle, then the observed brightness can be linked directly
to a distance (called luminosity distance $d_L$): the more distant the standard candle, the dimmer we see it. Astronomers tend to use not the
distance but its logarithm $\mu \sim \log_{10} d_L$, for historical reasons (supposedly because the eye
uses a roughly logarithmic scale to gauge brightness, apparently the magnitude system dates back to Greek astronomers). 
The magnitude-redshift diagram, a plot of $\mu$ versus
$z$, can then be used to study the way the universe has behaved over most of its lifetime.

Unfortunately not all supernovae are standard candles. There are two main classes that are due to very
different mechanisms. Stellar explosions of type Ia are thought to occur when a small dead star (a white dwarf)
exists in a binary system with a large, red star and begins to accrete material from that larger star. At some point
this cosmic cannibal begins to over-eat, and its stellar structure becomes unstable. At this point, reminiscent somewhat
of the unforgettable dinner scene in the Monty Python sketch, the white dwarf explodes in a gigantic conflagration.
Although the exact mechanism is not yet fully understood, it is plausible that such events always produce supernovae
of comparable luminosity, as the instability always occurs under about the same conditions. 

The second class is
due to supermassive stars running out of fuel: during most of the life of gigantic stars, the gravitational force of their
own mass that wants to crush them is balanced by the radiation pressure generated by burning hydrogen, and later
heavier elements. This is possible because it is energetically favorable to fuse light elements together, but once
the fusion process reaches lead, the energetics reverse and for heavier elements it is actually favorable to split
(this is why fusion plants would use light elements while fission plants use very heavy elements). So stars will eventually
run out of fuel, and thus will lose the radiation pressure. If the star is heavy enough then the matter itself also cannot
support the self-gravity of the star, and it will collapse under its own weight. The result is called a core-collapse
supernova, and much of the gravitational energy liberated in the collapse is radiated away in neutrinos and photons.
Such supernovae occur unfortunately with a wide variety of intrinsic luminosities and so are unsuitable for distance
measurements.

\begin{figure}[hbt]
\sidecaption
\includegraphics[scale=.45]{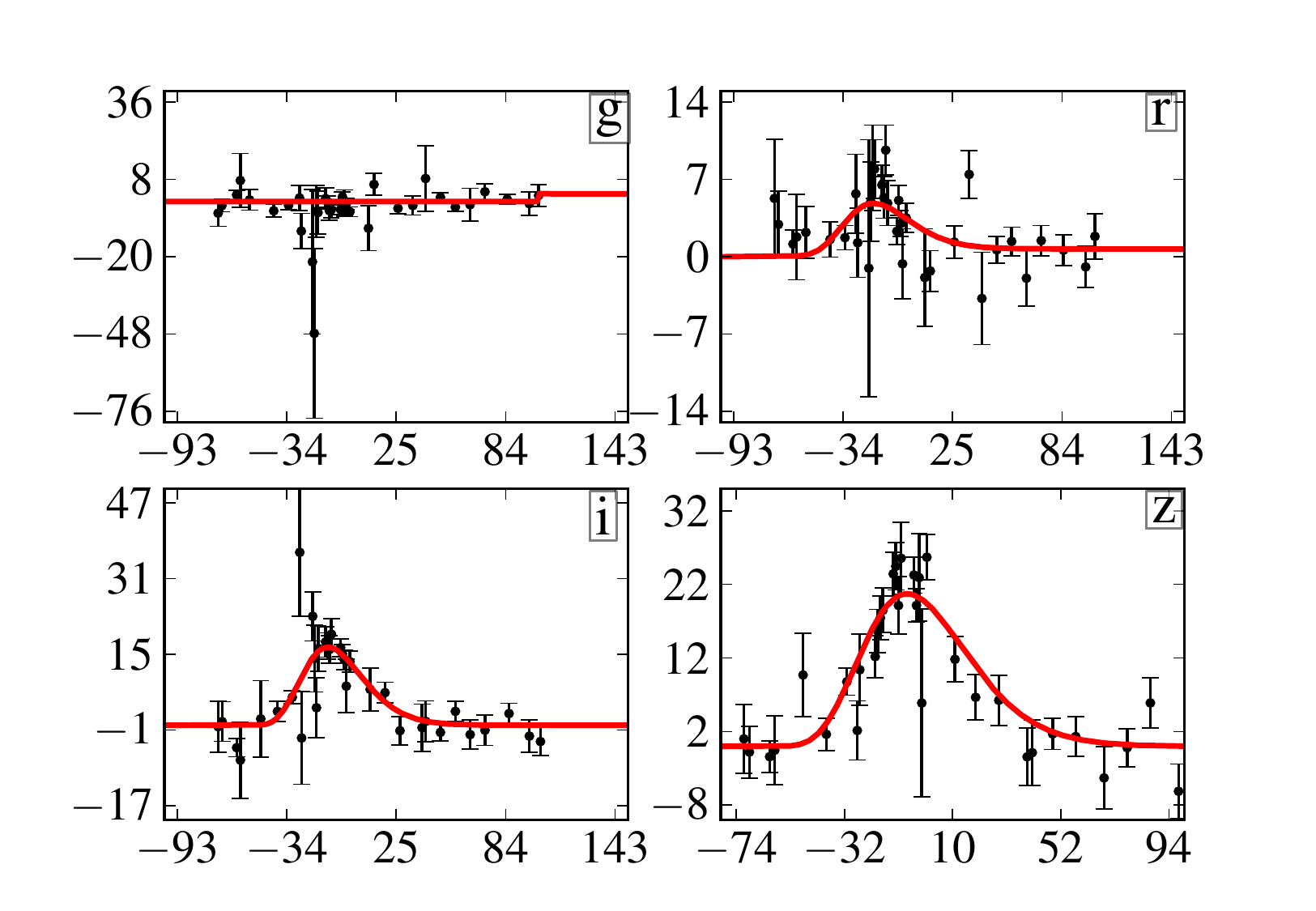}
\caption{\textbf{Fitting lightcurves to supernova data:} A simulated set of SN-Ia light curves in different bands
from the Supernova Photometric Classification Challenge \cite{Kessler:2010wk}, together with interpolating
curves from \cite{Newling:2010bp}.}
\label{fig:photo_sn}   
\end{figure}

Luckily the different types can be distinguished with the help of spectral analysis of the supernova light. But
measuring a spectrum requires much more light and effort than simply measuring the brightness of an object, as we
need to split the light into the different wavelengths. While taking spectra has been feasible for the hundreds
supernovae that have been observed to date, we are now seeing a transition where large surveys are finding thousands
of supernovae (e.g.\ the supernova data of the Sloan Digital Sky Survey, SDSS-II SN \cite{holtzman:etal2009, kessler:etal2009}), 
which are too many to take all the spectra. Future astronomical projects like the
large synoptic survey telescope (LSST \cite{lsst:2009pq}) will find tens of thousands to hundreds of thousands of supernovae
{\em per year}. It is impossible to follow up more than a tiny fraction of this data with spectroscopic observations.
The ``normal'' observations will still provide light-curves in several different color bands, of the kind shown in
Fig.~\ref{fig:photo_sn}. Such observations will yield some idea what kind of supernova we are looking at, but they
cannot provide the near-certainty of spectra. We are then faced with a stark choice: either
we throw away over 99\% of the data, or we develop a statistical method that is robust against mis-identification
of supernovae. Here we will make an attempt at providing such a method. The material presented here is
based on several publications \cite{beams_kunz,newling/etal:2011,Hlozek:2011wq} where more details can be found
(see also \cite{press,gong/etal:2010}).

In the following section, we introduce the BEAMS formalism and discuss in more detail the role of the probabilities.
We then present our choice of likelihood functions for the different types of supernovae in section \ref{sec:beamsn},
where we also provide some tests of the algorithm itself. In section \ref{sec:sdss} we apply the algorithm to the
SDSS-II supernova data. In the final section we summarize the chapter, providing conclusions and an outlook to
future work.

\section{Basic BEAMS}
\label{sec:toybeams}

\subsection{The BEAMS formalism}
\label{sec:formalism}
Let us first introduce the mathematical formalism (see also \cite{beams_kunz} for simple examples and basic tests).
We normally want to know
the posterior distribution $P(\theta|D)$ for parameters $\theta$ given data $D$. Now assume that there is an
additional dependence on the type of population that the data has been drawn from. For simplicity, let us assume
that there are two kinds of data, type $A$ (corresponding for example to type Ia supernovae) and type $B$
(all other kinds of supernovae). Introducing a type vector $\tau$ of the same length $N$ as the data vector $D$ and
with entries $\tau_i = A$ or $\tau_i = B$, we can then write
\begin{equation}
P(\theta| D) = \sum_{\tau} P(\theta,\tau|D) \label{eq:initial}
\end{equation}
where we marginalized over all $2^N$ possible $\tau$ vectors. It is obviously straightforward to generalize this
to an arbitrary number of different populations. The joint probability density $P(\theta,\tau|D)$ for a given
vector $\tau$ is probably difficult to determine directly, so we use Bayes theorem to rewrite it,
\begin{equation}
P(\theta,\tau|D)  = P(D|\theta,\tau) \frac{P(\theta,\tau)}{P(D)} .
\end{equation}
The ``evidence'' factor $P(D)$ is independent of both the
parameters and $\tau$ and is an overall normalization that can
be dropped for parameter estimation. We will further assume
here that $P(\theta,\tau) \approx P(\theta) P(\tau)$. This simplification
assumes that the actual parameters describing our universe are
not significantly correlated with the probability of a given
supernova to be of type Ia or of some other type. Although it is
possible that there is some influence, we can safely neglect it
for current data as our parameters are describing the large-scale
evolution of the universe, while the type of supernova should mainly
depend on local gastrophysics. In this case $P(\theta)$ is the usual
prior parameter probability. We will also assume $P(\tau)$ to separate into independent
factors,
\begin{equation}
P(\tau) = \prod_{\tau_i=A} P_i \prod_{\tau_j=B} (1-P_j) ,
\end{equation}
for a discussion of this approximation please see \cite{newling/etal:2011}.
Here the product over ``$\tau_i=A$'' should be interpreted 
as a product over those indices $i$ in the vector $\tau$ for which $\tau_i=A$. 
In other words, given
a population vector $\tau$ with entries ``$A$'' for SN-Ia and ``$B$'' for
other types, the total probability $P(\tau)$ is the product over all entries, 
with a factor $P_j$ if the j-th entry is ``$A$'' and $1-P_j$ otherwise (if the
j-th entry is ``$B$''). In this way $P_j$ is always a probability to find an entry
$\tau_j=A$ in the vector $\tau$ before using the data. Notice that we discuss here
only one given vector $\tau$, the uncertainty is taken care of by
the outer sum over all possible such vectors. The full expression
is therefore
\begin{equation}
P(\theta|D) \propto P(\theta) \sum_\tau P(D|\theta,\tau)  
\prod_{\tau_i=A} P_i \prod_{\tau_j=B} (1-P_j). \label{eq:posterior}
\end{equation}
The factor $P(D|\theta,\tau)$ is the likelihood, but now conditional
on the data types. This means when we write down later on an expression
for the likelihood, we can do it assuming that the type of each data point
is known.

The price to pay is that we then have to marginalize over all possible
vectors $\tau$, evaluating a sum composed
of $2^N$ terms for $N$ data points. The exponential scaling with
the number of data points means that we can in general not evaluate
the full posterior directly, but have to use a clever approximation. Here
we will instead make an additional assumption that the data points
are not correlated,
\begin{eqnarray}
P(D|\theta,\tau) &=& \prod_{i=1}^N P(D_i|\theta,\tau) \\
&=& \prod_{\tau_i=A} P(D_i|\theta,\tau_i=A)
\prod_{\tau_j=B} P(D_j|\theta,\tau_j=B) . \label{eq:like}
\end{eqnarray}
In the second line we have made the reasonable assumption that the
probability of observation $i$ does not depend on the assumed type
of the object $j\neq i$. We have also indicated that the likelihood of each
observation naturally splits into two populations, those which have entry
$A$ in $\tau$, and those with entry $B$. In general the form of these two
likelihood classes will be different. In toy model applications, we will usually
know how they look like, but for actual data they may be unknown and we
will have to leave some additional freedom.

The form of Eqs.~(\ref{eq:posterior}) and (\ref{eq:like}) allows for a huge computational simplification:
the posterior is the sum over all possible products of the type
$A_1 A_2 A_3 \ldots$, $B_1 A_2 A_3 \ldots$, $A_1 B_2 A_3 \ldots$, etc. 
This sum of $2^N$ terms can be generated simply by computing
the product over $N$ terms $\prod_k (A_k + B_k)$, and the posterior 
Eq.~(\ref{eq:posterior}) can be written as
\begin{equation}
P(\theta|D)\propto P(\theta)
\prod_{i=1}^N \left\{ P(D_i|\theta,\tau_i=A) P_i + P(D_i|\theta,\tau_i=B) \left(1-P_i \right) \right\} .
\label{beamsfull_algorithm}
\end{equation}
This is the form of the BEAMS posterior that we will be using for the rest of this chapter.

\subsection{BEAMS probabilities}
\label{sec:probabilities}

\begin{figure}[htbp!]
\begin{center}
$\begin{array}{@{\hspace{-0.1in}}c}
\includegraphics[width=0.7\columnwidth,trim = 0mm 0mm 0mm 12mm, clip]{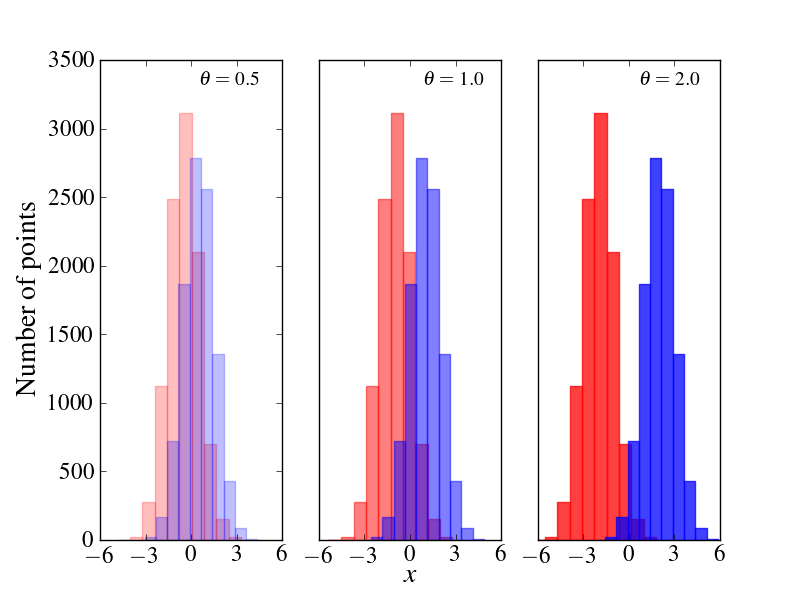}\\ [0.0cm]
\includegraphics[width=0.65\columnwidth]{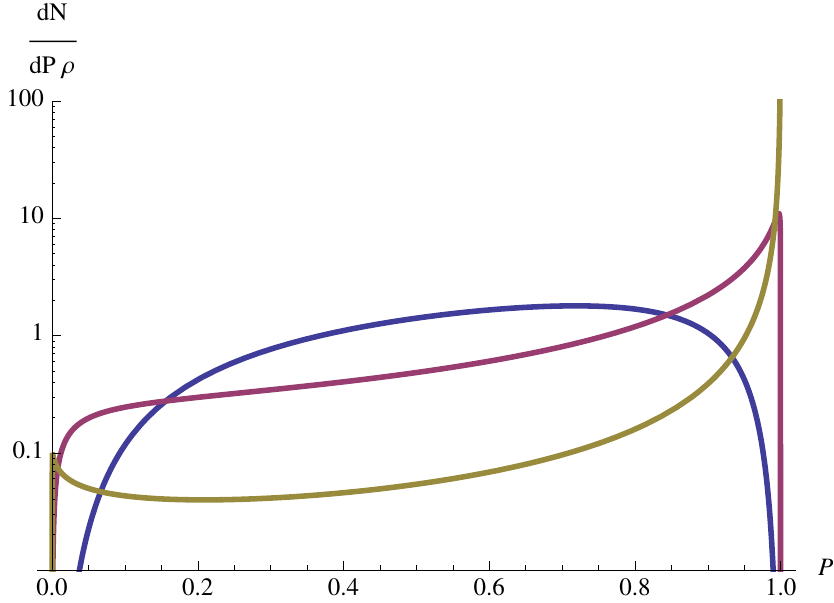}\\ [0.0cm]
\end{array}$
\caption{\textbf{Posterior probabilities:} the top panel provides an illustration of the two toy distributions, in the case of $\theta = 0.5, 1.0, 2.0$ (left to right). The bottom panel shows the probability histogram density plots, or number of red points with a given probability, where $dN^{(r)}(P)$ is given in Eq.~(\ref{eq:probdensity}) for $\theta=0.5$ (blue), $1$ (red) and $2$ (yellow). \label{fig:probtoy}}
\end{center}
\end{figure}

The probabilities in BEAMS are of central importance, so it is worth to take a closer look by studying a simple toy model: we assume that we are dealing with two populations (let us call them `red' and `blue') drawn from two normal distributions with means at $\pm \theta$ and equal variances of $\sigma^2=1$, see the top panel of Figure~\ref{fig:probtoy}. In addition to the basic formalism discussed in section \ref{sec:formalism} above, we will introduce an extra parameter $A$ that can adjust the relative normalization of the probabilities. As we will see, this parameter allows for automatically adjusting for an unknown relative rate of the two populations. We introduce the parameter as a change of the relative probability $B$ (the Bayes factor) to be of the red or blue kind:
\begin{equation}
B = \frac{P}{1-P} \rightarrow \tilde{B} = B A = A \frac{P}{1-P} = \frac{\tilde P}{1-\tilde{P}} \, .
\end{equation}
The effective, adjusted probability is then
\begin{equation}
\tilde P = \frac{A P}{1- P + A P} \, , \label{eq:aparam}
\end{equation}
and we will in the actual applications always use this probability and allow for a free $A$ that we will marginalize over.

%\vspace{-0.3in}
The equality of the variances of the two populations means that we are measuring the distance $\Delta=2\theta$ between the two mean values in units of the standard deviation. We also allow for different numbers of points drawn from the red and blue Gaussians through a `rate parameter' $\rho\in [0,1]$ that gives the probability to draw a red point. If we draw $N$ points in total, we will then have on average $\rho N$ red points and $(1-\rho)N$ blue points. The likelihood for a set of points $\{x_j\}$, with $j$ running from $1$ to $N$, is then
\begin{equation}
P(\{x_j\}|\theta) = \prod_{j=1}^N \frac{1}{\sqrt{2\pi}} \left(P e^{-\frac{1}{2}(\theta-x_j)^2} + (1-P) e^{-\frac{1}{2}(\theta+x_j)^2} \right) 
\end{equation}
for $P=\rho$.

To simplify the analysis we assume that we are dealing with large samples so that $\theta$ is determined to high precision, with an error much smaller than $\sigma$. In this case (and since this is a toy model) we can take the parameter $\theta$ fixed. We also note that if we are running this in BEAMS with a true prior probability $P=\rho$ then we would find a normalization parameter $A=1$, while for $P=1/2$ we would obtain $A=\rho/(1-\rho)$, and we again assume that this parameter can be fixed to its true value. Then it is easy to see that if we leave the probability for point $i$, $P_i$, free, we find a Bayes factor
\begin{equation}
B=\frac{P(\{x_j\}|\theta, P_i=1)}{P(\{x_j\}|\theta,P_i=0)} = \frac{e^{-\frac{1}{2}(\theta-x_i)^2}}{e^{-\frac{1}{2}(\theta+x_i)^2}}
= e^{2\theta x_i} .
\end{equation}
In other words, $\ln(B) = x_i \Delta$, just the value of the data point times the separation of the means. If the point is exactly in between the two distributions, $x_i=0$, then $B=1$, i.e. its BEAMS posterior probability to be red or blue is equal. This means that if we want to think of the BEAMS posterior probability as the probability to be red or blue, we should update the Bayes factor with $A$, i.e. use $\tilde{B} = B A$, with an associated probability $P=\tilde{B}/(1+\tilde{B})$. We also see that the probability to be red increases exponentially as $x_i$ increases. As we will see below, this reflects the fact that the number of red points relative to the blue points increases in the same way. The rapidity of this increase is governed by the separation, $\Delta,$ of the two distributions.

What is the distribution of the posterior probabilities, i.e. the histogram of probability values, and what determines how well BEAMS does as a typer in this example? The number of red points in an interval $[x,x+dx]$ is just given by the `red' probability distribution function at this value, times $dx$. To plot this function in terms of $P$ we also need 
\begin{eqnarray} x(P)&=&\frac{\ln(B)}{\Delta}=\frac{\ln(P/(1-P))}{\Delta}\, ,\\ ~~~\frac{dP}{dx} &=& {\Delta P}{(1-P)} \, .
\end{eqnarray} 
The probability histograms for the red (r) and blue (b) points, normalized to $\rho$ and $1-\rho$ respectively, then are:
\begin{eqnarray}
 \label{eq:probdensity}
dN^{(r)}(P) &=& \frac{\rho}{\sqrt{2\pi}\Delta}  \frac{dP}{P(1-P)}  \exp\left\{ -\frac{1}{2} \left( \frac{\ln[P/(1-P)]}{\Delta} - \theta \right)^2 \right\} \, ,\\
\qquad dN^{(b)}(P) &=& \frac{1-\rho}{\sqrt{2\pi}\Delta} \frac{dP}{P(1-P)}  \exp\left\{ -\frac{1}{2} \left( \frac{\ln[P/(1-P)]}{\Delta} + \theta \right)^2 \right\} \, .
\end{eqnarray}
We plot $dN^{(r)}/dP/\rho$ for $\theta=0.5$, $1$ and $2$ in the lower panel of Figure~\ref{fig:probtoy}. We see how the values become more concentrated around $P=1$ for larger separation of the distributions, i.e. BEAMS becomes a ``better'' typer. But for very large separations there are also suddenly more supernovae at low $P$ (yellow curve). The reason is that BEAMS does not try to be the best possible typer, instead it respects the condition that the probabilities have to be unbiased, in the sense that
\begin{equation}
\frac{dN^{(r)}}{dN^{(b)}} = \left( \frac{P}{1-P} \right) \left( \frac{\rho}{1-\rho} \right) = B A = \tilde{B} . \label{eq:nonbias}
\end{equation}
Since BEAMS only uses the information coming from the distribution of the values, its power, as reflected in the distribution of probability values $dN(P)$, is given by how strongly the distributions are separated. If they are identical ($\theta=0$) then BEAMS can only return $P=1/2$ while for larger $\theta$ there is a stronger preference for one type over another. But given the two populations, we can in principle derive the probability histogram by just looking at the ratio of data points of either type at each point in data space, there is nothing else BEAMS can do. Also, in order for the probabilities to be unbiased (up to the rates which are taken into account by $A$) if there are, say, 200 red points in the $P=0.9$ bin and only 10 in the $P=0.8$ bin, then we need to find about two blue points in the $P=0.8$ bin, but 20 in the $P=0.9$ bin. Although this looks like a significant misclassification problem, it is just a reflection of Eq.~(\ref{eq:nonbias}) and is actually the desired behavior: BEAMS is not a classification algorithm (see e.g.~\cite{Kessler:2010wk,Kessler:2010qj,Newling:2010bp} for efforts in that direction) but instead a way to compute the posterior pdf of the parameters $\theta$. The property of unbiased probabilities is required to get unbiased parameter constraints, and indeed for that purpose we never classify any data points. Instead we leave them in a superposition of different types, weighted by the associated probabilities as encoded in the marginalization over $\tau$ in Eq.~(\ref{eq:initial}).

\section{Application of BEAMS to supernova observations}\label{sec:beamsn}

In this section we will complete first our discussion of the posterior (\ref{beamsfull_algorithm}) by providing explicit
expressions for the two likelihood functions. We will say a few words on the numerical strategy used to explore
the posterior parameter distribution and check the performance of the algorithm with a range of tests. This section
and the next is based on the results obtained in \cite{Hlozek:2011wq}.

Before entering the likelihood discussion, we would like to remind the reader that supernova data is given
in the form of a distance modulus $\mu$ as a function of redshift $z$. In addition, the distance modulus depends
on a set of cosmological and nuisance  parameters $\theta$. The cosmological parameters $\{H_0,\Omega_m, \Omega_\Lambda\}$ are the true quantities of interest for us. Here $H_0$ is the expansion rate of the universe today
(the Hubble constant), and
the $\Omega_j$ are the relative energy densities in matter $m$ and a cosmological constant $\Lambda$.
See for example the book by Scott Dodelson \cite{Dodelson:2003ft} for a good introduction to cosmology.

The distance modulus is related to the cosmological model via:
\begin{equation}
\mu(z, \theta) = 5\log d_L(z,\theta) + 25 \label{eq:mu}\,,
\end{equation}
where
\begin{equation}
d_L(z, \theta) = \frac{c(1+z)}{\sqrt{\Omega_k}H_0} \sinh\left(\sqrt{\Omega_k}\int_0^z \frac{dz}{E(z)}\right) \label{eq:dl}
\end{equation}
is the luminosity distance measured in Megaparsec (Mpc), and the normalized expansion rate is given by 
\begin{equation}
E(z) \equiv \frac{H(z)}{H_0} = \sqrt{\Omega_m(1+z)^3 + \Omega_k(1+z)^2 + \Omega_{\Lambda}}.
\end{equation}
The relative energy densities of matter ($\Omega_m$), curvature ($\Omega_k$) and the cosmological constant ($\Omega_{\Lambda}$) obey the relation $\Omega_m+\Omega_k + \Omega_{\Lambda}=1$, which we use to express
$\Omega_k$ in terms of the other $\Omega$'s. Notice that $\Omega_k<0$ is possible, in which case $\sqrt{\Omega_k}$ in Eq.~(\ref{eq:dl}) becomes imaginary and the hyperbolic sine becomes a normal sine function instead -- the limit $\Omega_k=0$ is also well defined.
The distance modulus is defined as the difference between the absolute and apparent magnitudes of the supernova, $\mu = m-M,$ with additional corrections made to the apparent magnitude for the correlations between brightness, color and stretch and a K-correction term related to the difference between the observer and rest-frame filters, for example. The corrections are typically made within the model employed in a light-curve fitter, such as that for MLCS2k2. 

In this application of BEAMS we have assumed that the distance modulus $\mu$ is obtained directly from the light-curve fitter (such as is the case for fitters which use the MLCS2k2 light-curve model), however this is not an implicit assumption. In the case of the SALT light-curve fitter, the distance modulus would be reconstructed using a framework such as that outlined in \cite{marriner/etal:2011} before including in the BEAMS algorithm. We will also always assume that the distance modulus has been obtained {\em under the assumption that all supernovae are of type Ia}. This means that it is straightforward to write down the likelihood for type-Ia supernovae, but that we need to do extra work for the non-Ia supernovae. It would of course be preferable to have distance moduli for all possible supernova types, but this is still an active research topic in astronomy.

\subsection{Choice of likelihood}

\subsubsection{Likelihood of type-Ia supernovae}

Following the standard astronomical literature, the Ia likelihood is modeled as a Gaussian probability distribution function (pdf) for the observed distance modulus $\mu_i$ centered around the theoretical value $\mu(z,\theta)$ with a variance $\sigma_{\mathrm{tot},i}^2$:
\begin{equation}
P(\mu_i|{\theta}, \tau_i = 1) = \frac{1}{\sqrt{2\pi}\sigma_{\mathrm{tot},i}}\exp\left(-\frac{(\mu_i - \mu(z_i,{\theta}))^2}{2 \sigma_{\mathrm{tot},i}^2}\right).\label{ialike}
\end{equation}

Again following standard practice, we model the error on the distance modulus of each supernova as a sum in quadrature of several independent contributions,
\begin{equation}
\sigma_{\mathrm{tot},i}^2 = \sigma_{\mu,i}^2 + \sigma_{\tau}^2 + \sigma_{\mu,z}^2, \label{eq:pecvel} 
\end{equation}
where $\sigma_{\mu,i}$ is the error obtained from fits to the SN light-curve, $\sigma_{\tau}$ is the characteristic intrinsic dispersion of the supernova population, which we add as an additional global parameter to the vector $\theta$ with Jeffreys' prior. The constraints do not depend strongly on the prior used for the intrinsic dispersion. The error term $\sigma_{\mu,z}$ converts the uncertainty in redshift due to measurement errors and peculiar velocities into an error in the distance of the supernova as:
\begin{equation}
\sigma_{\mu,z} = \frac{5}{\ln(10)}\frac{1+z}{z(1+z/2)}\sqrt{\sigma_z^2 + (v_{pec}/c)^2}, \label{error}
\end{equation}
with $\sigma_z$ as redshift error, and $v_{pec}$ as the typical amplitude of the peculiar velocity of the supernova, which we take as $300~\mathrm{km}\mathrm{s}^{-1}$ \cite{lampeitl:etal2009,kessler:etal2009}.

\subsubsection{Likelihood of all other supernovae}\label{section:nonIalike}

\begin{figure}[htbp!]
%\vspace{-0.5in}
\begin{center}
\includegraphics[width=\columnwidth,trim = 0mm 0mm 10mm 10mm, clip]{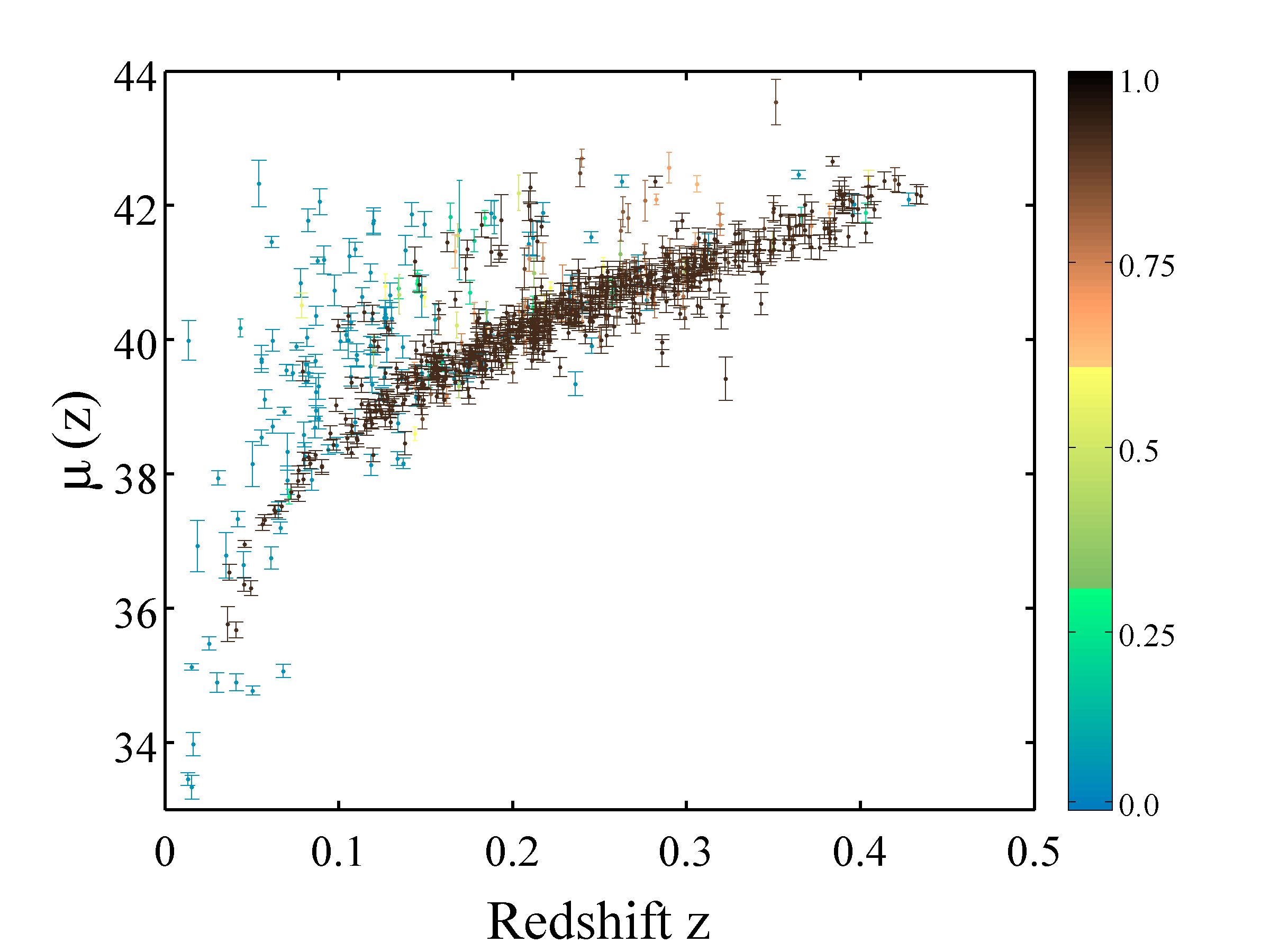}\\[0.0cm]%-1.2cm]
\caption{\textbf{SDSS-II SN data:} the photometric sample of the full three seasons of SDSS-II SN survey. The 792 points are all those with host galaxy spectroscopic redshifts. The sample includes 297 spectroscopically confirmed SNe, and the data points are color coded using probabilities from the PSNID typer \cite{masao_typer, sako/etal2011_typer2} from low (blue) to high (dark brown).\label{fig:level|II}}\end{center}
\end{figure}

The general form of the non-SNIa likelihood will be complicated since there are several sub-populations. Given the limited number of non-SNIa in the SDSS-II SN data set however, (see Figure~\ref{fig:level|II}) we will model it with a single mean and a dispersion. If one chooses to describe a population using only a mean and a variance, statistically the least-informative (maximum entropy) choice of pdf in this case is also a Gaussian \cite{jaynes}, 
\begin{equation}
P(\mu_i|{\theta}, \tau_i = 0) = \frac{1}{\sqrt{2\pi}s_{\mathrm{tot},i}}\exp\left(-\frac{(\mu_i -  \eta(z_i,{\theta}))^2}{2 s_{\mathrm{tot},i}^2}\right) \label{nonialike} .
\end{equation}
As we do not know the mean $\eta$ and variance $s_{\mathrm{tot},i}^2$ of the non-Ia population, we describe them with additional parameters.  
We will keep the parametrization of the mean very general (see below) but for the variance we restrict ourselves to the same form as for the Type Ia supernovae, Eq.~(\ref{eq:pecvel}), but with a potentially different intrinsic dispersion $s_\tau^2$ described by an independent parameter (again with a Jeffreys' prior). We assume that the measurement errors and the contribution from the peculiar velocities enter in the same way for Type Ia and other supernovae and so keep these terms identical.

We do not know what to expect for the mean of the non-Ia pdf and so we allow for a range of possibilities. As the brightness is linked to the luminosity distance through Eq.~(\ref{eq:mu}), we describe the expected non-Ia distance modulus (as provided by the light-curve fitter which assumes actually a type-Ia supernova) as a deviation from the theoretical value, $\eta(z,{\theta})=\mu(z,{\theta})+\Upsilon(z)$, where we consider the following Taylor expansions of the difference as a function of redshift: %\vspace{-0.35in}
\begin{equation}
\Upsilon(z) = \eta(z,{\theta})-\mu(z,{\theta}) \propto \sum^3_{i=0} (a_i z^i)/(1+dz) .
\label{nonIadist}
\end{equation}
We consider the cases where we set different combinations of the parameters $(a_i, d)$, to zero, and employ a criterion based on model probability to decide which of these functions to use. We note that the explicit link of $\eta(z,{\theta})$ to $\mu(z,{\theta})$ carries a risk that the non-Ia likelihood can influence the posterior estimation of the cosmological parameters. For this reason we verify that the contours do not shift when we set directly  $\eta(z,{\theta})=\Upsilon(z)$, although we will need a higher-order expansion in general (and of course the recovered parameters of the function $\Upsilon(z)$ will change). In general, as long as the basis assumed has enough freedom to fit the deviation in the distance modulus of the non-Ia population from the Ia model, the inferred cosmology will not be biased.

For a cosmological analysis we just marginalize over the values of the parameters in $\Upsilon(z)$, but these parameters contain information on the distribution of non-Ia type SN and thus their posterior is of interest as well, allowing us to gain insight into the distribution characteristics on the non-Ia population at no additional `cost'. 

The simple binomial case considered here, where the non-Ia population consists of all types of core-collapse SNe, is probably too simplistic to accurately describe the distribution of non-Ia supernovae. In general one could include multiple populations, one for each supernova type, which would yield a sum of Gaussian terms in the full posterior. In addition, the forms describing the distance modulus of the non-Ia population are chosen to minimize the cosmological information from the non-Ia's (we always test for a deviation from the cosmological distance modulus), however, the parameterization of the non-Ia distance modulus could be improved by investigating the distance modulus residuals from simulations, as the major contributions to the distance modulus residuals appear to be the core-collapse luminosity functions,  along with the specific survey selection criteria and limiting magnitude, see \cite{falck:dist}.  While current SN samples do not include a large enough sample of non-Ia data to test for this, larger data sets (such as the data from the BOSS SN survey) will allow for a detailed analysis of the number (and form) of distributions describing the contaminant population.

\subsection{Numerical methodology}

In this work, the BEAMS algorithm is implemented within a Markov Chain Monte Carlo (MCMC) framework, and the Metropolis-Hastings \cite{metropolis-hastings} acceptance criterion was used. 
We use the cosmological parameters $\{\Omega_m, \Omega_\Lambda, H_0\}$ in the case of the $\chi^2$ approach on the \textit{spectro} and \textit{cut} samples described below, and add additional parameters $\{A, \sigma_\tau, s_\tau, \vec{a}\}$   in the case of the BEAMS application. The parameters $\vec{a} = \{a^0, a^1, a^2\}; d = a^3 = 0$ are for the quadratic model, in the other models for $\Upsilon(z)$ we adjust the parameters accordingly. The chains were in general run for around 100 000 steps per model; this was sufficient to ensure convergence. We test for convergence using the techniques described in \cite{dunkley/etal:2005}. We impose positivity priors on the energy densities of matter and dark energy, and impose a flat prior on the Hubble parameter between $20 <  H_0 <  100$~kms$^{-1}$Mpc$^{-1}$. The Hubble parameter is marginalized over given that we do not know the intrinsic brightness of the supernovae, but through the distance modulus are only sensitive to the \textit{relative} brightness of the supernovae. We impose broad Gaussian priors on the parameters of the non-Ia likelihood function, and step logarithmically in the probability normalization parameter $A$, as well as the intrinsic dispersion parameters of both the Ia and non-Ia. distributions.

\subsection{Comparison to standard $\chi^2$ methods}
\label{cutssection}
The primary difference between BEAMS and current methods is that the latter either require that all data are spectroscopically confirmed, or apply a range of quality cuts based on selection criteria. Here we will compare the performance of BEAMS to these two approaches, by processing the data that pass the required selection criteria using the Ia likelihood, Eq.~(\ref{ialike}). We will hereafter refer to this as the $\chi^2$ approach.\\\\
We use the following samples\footnote{We apologize for the use of technical jargon in the description of the samples.}:
\begin{itemize}
\item{\textbf{spectro sample}}:\\The sample containing only spectroscopically confirmed supernovae. In addition to spectroscopic confirmation we will also apply a cut on the goodness-of-fit probability from the light-curve templates within the MLCS2k2 model, $P_{\mathrm{fit}}> 0.01$, and a cut on the light-curve fitter parameter $\Delta >  -0.4,$ where $\Delta$ is a parameter in the MLCS2k2 model describing the light-curve width-luminosity correlation. MLCS2k2 was trained using the range $-0.4 < \Delta < 1.7$ \cite{jha:etalMLCS2k2}, hence we restrict the sample to $\Delta >  -0.4,$ which is a cut typical in current SN surveys, and so we introduce the cut to provide comparison between datasets. We process this \textit{spectro} sample using the $\chi^2$ approach.

\item{\textbf{{cut} sample}}:\\This larger sample is selected both by removing $5\sigma$ outliers from a moving average fit to the Hubble diagram including both photometric and spectroscopically confirmed data and applying a cut to the sample, including only data with a high enough probability, $P_{\mathrm{typer}}>0.9$ (where the probability comes from a general supernova typing procedure, such as PSNID, described in \cite{masao_typer, sako/etal2011_typer2}). We choose to use the PSNID probabilities to make the probability cut on the sample ($P_{\mathrm{typer}}>0.9$); if the MLCS2k2 probabilities had themselves been used to make a \textit{cut} sample, then objects would only be included if they had probabilities greater than, for example, $P_{\mathrm{fit}} > 0.9$ In addition, we impose a cut on the goodness-of-fit of the light-curve data to the Type Ia typer, $\chi^2_{lc} < 1.8,$ a cut on the goodness-of-fit probability from the light-curve templates within the MLCS2k2 model, $P_{\mathrm{fit}}> 0.01$, and a cut on $\Delta >  -0.4.$ In this \textit{cut} sample case we then use standard the $\chi^2$ cosmological fitting procedure on the sample, and so set the Ia probability of all points to one.

\item{\textbf{photo sample}}:\\This sample is the one to which BEAMS will be applied, and will include all the photometric data with host galaxy redshifts. As in the previous two cases, we include only data which have $P_\mathrm{fit} > 0.01, \Delta > -0.4$. 
\end{itemize}
Note that the \textit{spectro} sample will be included in all the three samples described above.  While the \textit{spectro} and \textit{cut} samples have by definition $P_\mathrm{Ia} = 1$ (as they are analyzed in the $\chi^2$ approach), we do not set the probabilities to unity when applying BEAMS to the full sample -  the \textit{spectro} subsample within the larger \textit{photo} sample will be treated `blindly' by BEAMS. The \textit{spectro} sample is the one most similar to current cosmological samples, and will be used to check for consistency in the derived parameters between BEAMS applied to the photo sample and the $\chi^2$ approach on the \textit{spectro} sample.

\subsection{Tests on simulated data}

To test the BEAMS algorithm explicitly we need a completely controlled sample, where all variables (such as the non-Ia model and the SN-Ia probabilities) are directly known and where we can verify that the algorithm is able to recover them correctly. In addition, we use this data set to check that we recover the correct shape of the non-Ia distance modulus $\eta(z)$ since the true $\eta(z, \vec{\theta})$ is known for this sample only.
We simulate a population of 50 000 SNe, with redshifts drawn from a Gaussian distribution, $z \sim \mathcal{N}(0.3,0.15),$ and distance moduli drawn from a flat $\Lambda$CDM universe with $(\Omega_m, \Omega_\Lambda, H_0) = (0.3, 0.7, 70)$. The non-Ia population includes a contribution to the distance modulus, $\eta(z,\vec{\theta}) = \mu(z,\vec{\theta}) +  a^0+ a^1z + a^2z^2  $, where we choose $(a^0, a^1,a^2) = (1.5,1,-3).$ We assign $P_{Ia}$ probabilities from a model $dN/dP_{Ia} = A_1P_{Ia} +A_2P_{Ia}^2$; with $A1 = -0.9$; $A2 = 1.9$. We then assign the \textit{types} from the two samples (of Ia's and non-Ia's), i.e. we choose a random number $t$ and if $t \leq P_{Ia}$ (i.e. the type also follows the same relationship as the probability) we take the data point to be a Ia, and if $t > P_{Ia}$ we assign it as a non-Ia, until we run out of data points from either sample. This procedure reduces the sample size from 50 000 to 37529, but guarantees unbiased probabilities. 

\begin{figure}[htbp!]
\begin{center}
$\begin{array}{@{\hspace{-0.25in}}l}
\includegraphics[width=\columnwidth,trim = 0mm 0mm 10mm 10mm, clip]{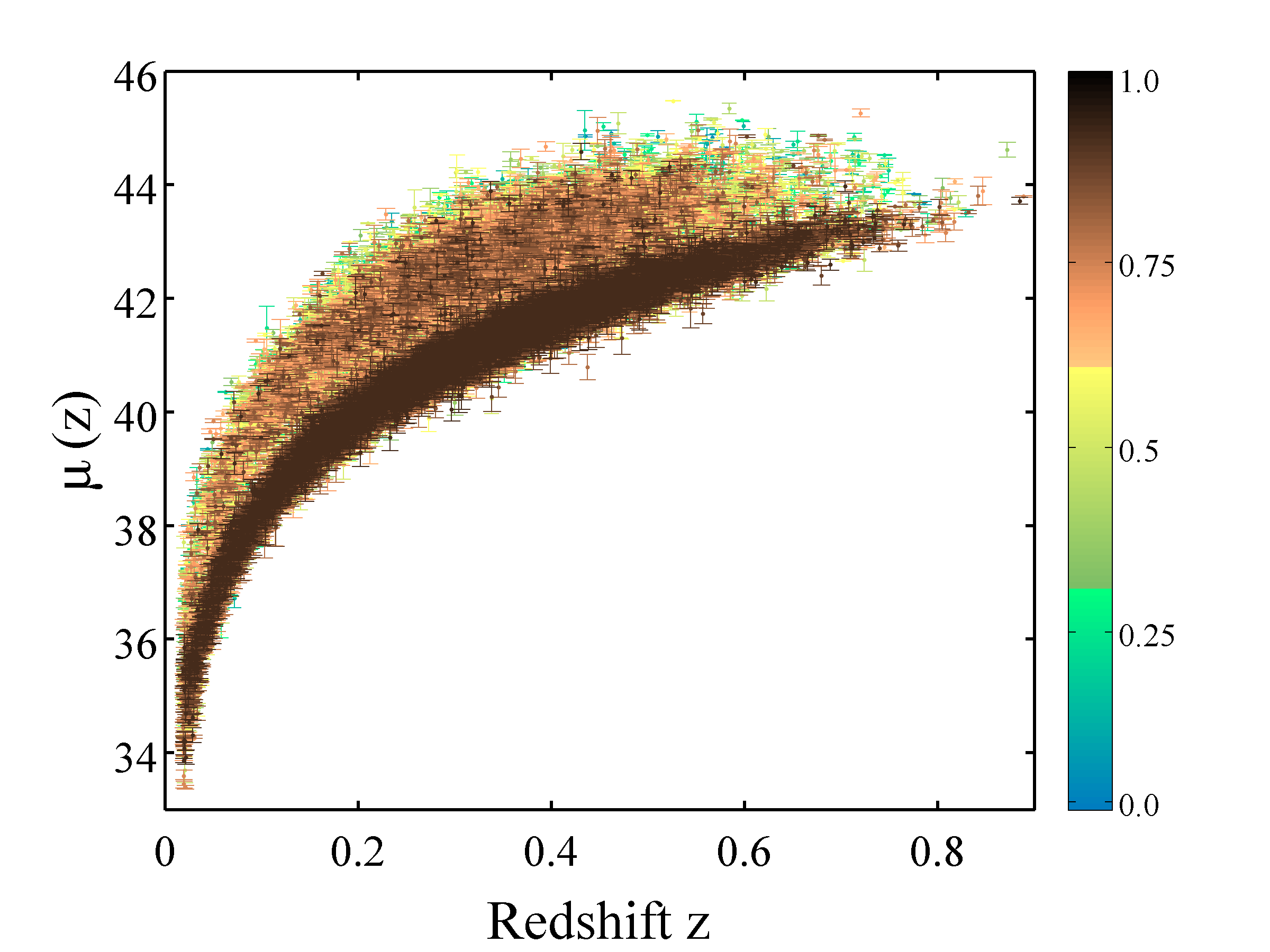}\\[0.0cm]
%,trim = 0mm 40mm 0mm 60mm, clip
 \end{array}$
 \caption{\textbf{Gaussian data:} 37529 points simulated according to a Gaussian distributions around a distance modulus in a flat $\Lambda$CDM model for the Ia population (25000 points) and with extra terms up to quadratic order in redshift for the non-Ia population. The points are colored according to their simulated probabilities from blue (low probability) to dark brown (high probability).  \label{fig:levelI}}
\end{center}
 \end{figure}
We assign a `measurement error' to each distance modulus of $\sigma_\mu = 0.1;$ add an intrinsic error $ \sigma_{\tau} = 0.16$ and a peculiar velocity error based on Eq.~(\ref{eq:pecvel}), with $v_{pec} = 300 \mathrm{kms}^{-1}$. We then randomly scatter the data points based on the total errorbar. To mimic what happens in a light-curve fitter, only the measurement error is recorded, however. When performing parameter estimation on the points we either add this measurement error in quadrature to the other terms whose amplitudes are fixed (in the case of the $\chi^2$ approach), or we estimate the magnitudes of the intrinsic dispersion when we apply the BEAMS algorithm. We randomly choose $10\%$ of the Ia data and assign \textit{spectro} status; this represents the data that are followed up by large telescopes on the ground. This \textit{spectro} sample is drawn so that we can compare the BEAMS-estimated result to the $\chi^2$ approach on a smaller sample. The data are shown in Figure~\ref{fig:levelI}. In the BEAMS analysis we checked on a small number of simulated samples that the results obtained were unbiased - a full Monte Carlo simulation of bias is beyond the scope of this work.

\subsubsection{Performance on cosmological parameters}

\begin{figure}[htbp!]
\begin{center}
$\begin{array}{@{\hspace{-0.25in}}l}
\includegraphics[ scale = 0.45, trim = 0mm 60mm 0mm 60mm, clip]{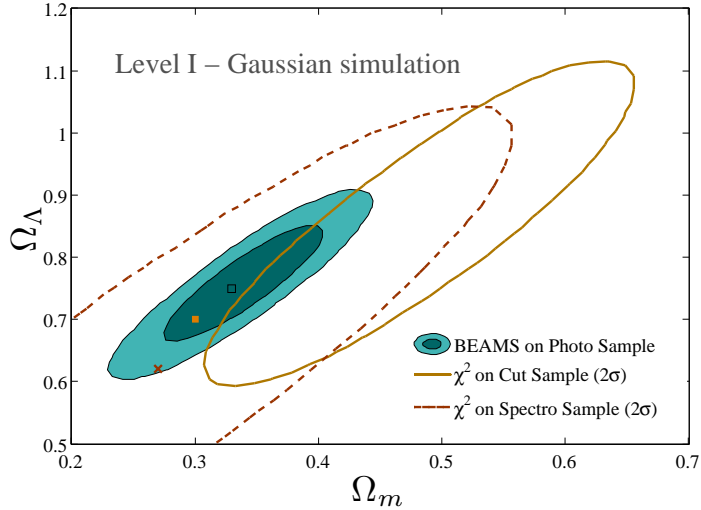}\\[0.0cm]
%,trim = 0mm 40mm 0mm 60mm, clip
 \end{array}$
 \caption{\textbf{Analysis of Gaussian simulation:} We show the $2\sigma$ contours in the $\Omega_m, \Omega_\Lambda$ plane for the simulated Gaussian data. The BEAMS constraints (filled contours with best fit indicated by the black open square) are consistent with the input cosmology (brown square), as is the `spectroscopic' sample (dashed contours, best-fit indicated by brown cross). The `cut' sample on the other hand is biased by over $2\sigma$ in spite of a relatively stringent cut on probability of $P_{\mathrm{cut}} = 0.9$; stronger cuts will recover the true cosmology at the cost of sample size.  \label{fig:contourI}}
\end{center}
 \end{figure}

We show in Fig.~\ref{fig:contourI} the $2\sigma$ confidence contours in the $\Omega_m, \Omega_\Lambda$ plane when analyzing the Gaussian simulation with BEAMS (filled contours), the `spectroscopic' sample (dashed contours) and the `cut' sample (solid contours). We see that both BEAMS and the spectroscopic sample are consistent with the input cosmology (filled brown square at $\Omega_m=0.3$, $\Omega_\Lambda=0.7$), but BEAMS can use the additional information in the data and is able to provide much tighter constraints. The cut sample has also smaller contours than the spectroscopic sample (although larger than BEAMS) but is biased with respect to the input cosmology.

\begin{figure}[htbp!]
\sidecaption
\includegraphics[width =0.63\textwidth, trim = 0mm 0mm 0mm 10mm, clip]{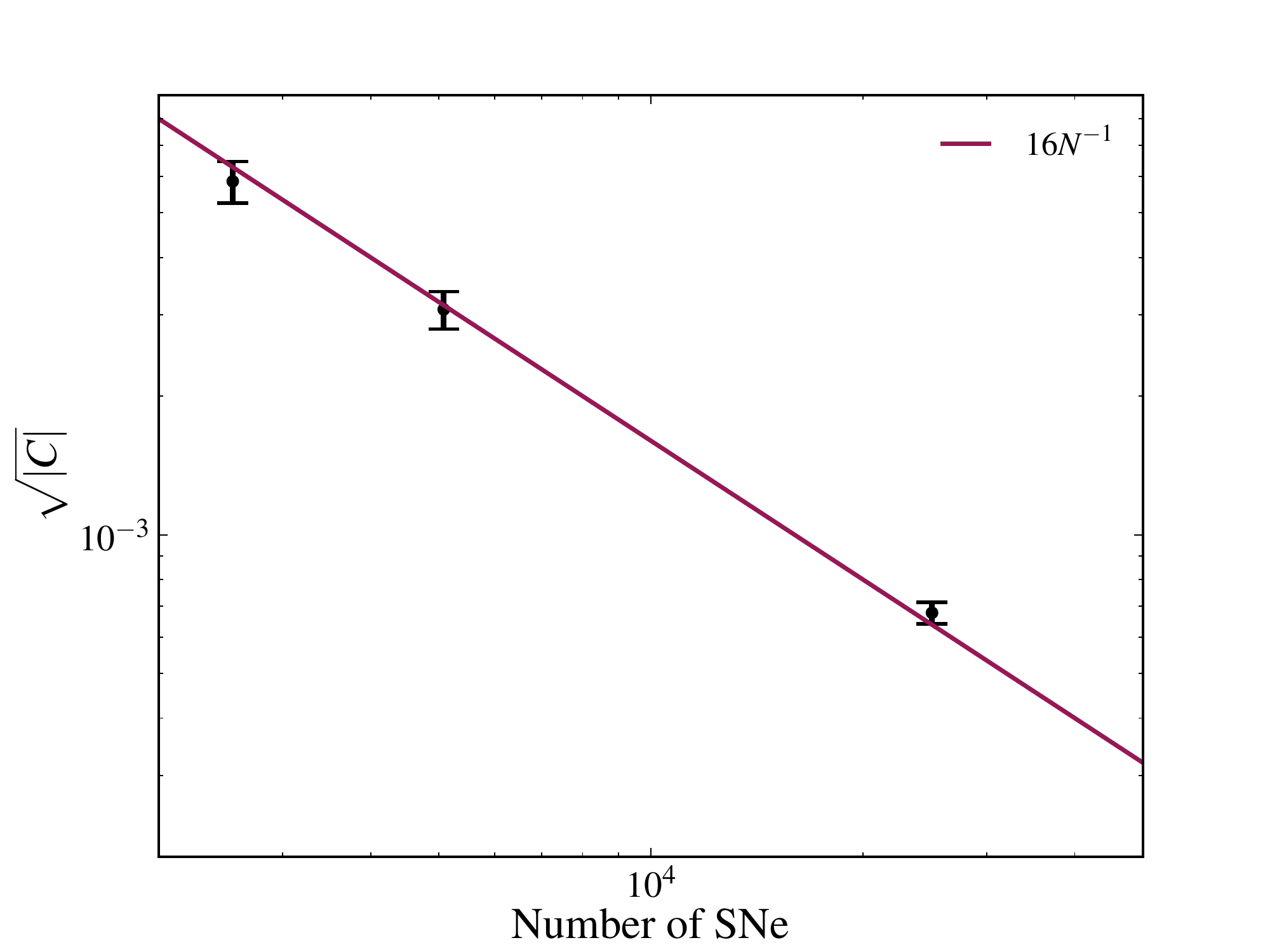}
\caption{{\bf{Errors scale with number of SNe:}} the size of the error ellipse, approximated by the square root of the determinant of the two-dimensional chain of $\Omega_m, \Omega_\Lambda$ shows the reduction in size with increasing the number of SNe in the simulation.  \label{fig:scaling}} 
 \end{figure}

As discussed in \cite{beams_kunz} for the one-dimensional case, the effective number of SNe that result when applying BEAMS scales as the number of spectroscopic SNe and the average probability of the dataset multiplied by the remainder of the photometric sample, $\sigma \rightarrow \sigma/\sqrt{N_\mathrm{spec} + \langle P_\mathrm{Ia} \rangle N_\mathrm{photo}}.$ In the two-dimensional case, the square root would be removed as the area of the ellipse scales with the increase in the effective number of supernovae.
In our applications we have, however, not used the fact that we know that some points are confirmed as Type Ia. In other words, the probability of each data point was taken from the light-curve fitter and was not adjusted to one or zero depending on the known type. Hence we expect the size of the contours in the $i-j$ plane to scale as 
\begin{equation} 
C_{ij}^{1/2} \rightarrow\frac{C_{ij}^{1/2}}{\langle P_\mathrm{Ia}\rangle N_\mathrm{photo}} \label{eq:error size} 
\end{equation}
We compute the size of the error ellipse for various Gaussian simulations as a function of the size of the simulation, shown in Figure~\ref{fig:scaling}, for one particular model of the probabilities, and hence one value of $\langle P \rangle$. We impose a prior on the densities, and hence the ellipses are not closed for smaller samples. For large enough sample sizes the ellipse is closed and we observe that the error ellipses scale in area as $\propto 1/\langle P_{\mathrm{Ia}}\rangle N,$ which is consistent with earlier results \cite{beams_kunz}. In general then, one would obtain a different constant factor $\langle P \rangle$ in Figure~\ref{fig:scaling} for different simulated probability distributions.

\subsubsection{Constraining $\Upsilon(z)$ forms for the non-Ia population}

The Gaussian simulation described in this section uses a quadratic model for the differences between the standard $\Lambda$CDM $\mu(z)$ and the non-Ia distance modulus.  We test here that assuming a different functional form while performing parameter estimation does not significantly bias the inferred cosmology. We define the effective $\chi^2$ as $-2\ln\mathcal{L},$ where the posterior $\mathcal{L}$ is given by Equation~(\ref{beamsfull_algorithm}), and we provide values relative to the simplest linear model for $\Upsilon(z)$. The goodness-of-fit of the distributions to the data is summarized in Table~\ref{table:level1dist}.
 \begin{figure}[htbp!]
\sidecaption[c]
\vspace{-0.6in}
\includegraphics[width=0.63\textwidth, trim = 0mm 45mm 0mm 50mm, clip]{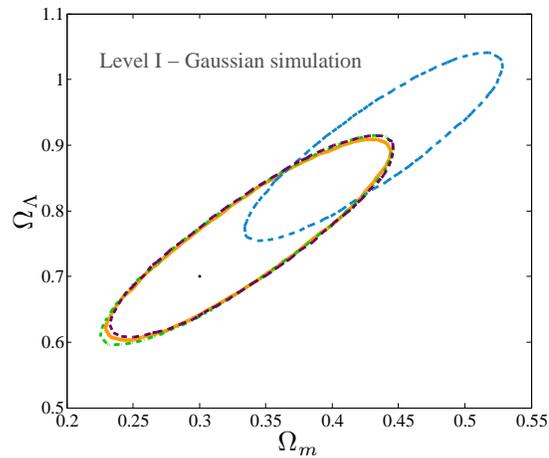} \\[2cm]
\caption{\textbf{Different $\Upsilon(z)$ for the non-Ia likelihoods}: $2\sigma$ constraints in the $\Omega_m,\Omega_\Lambda$ plane for different versions of the non-Ia distance modulus function, for the Gaussian simulation. We simulated a quadratic model, and ran BEAMS assuming a linear, quadratic, cubic and Pad\'{e} form for $\Upsilon(z)$, as described in Section \ref{section:nonIalike}. As expected, the linear model does not have enough freedom to capture the non-Ia distribution. \label{fig:dist}} 
 \end{figure}
In Figure~\ref{fig:dist} we show that BEAMS is reasonably insensitive to the assumed form of the non-Ia likelihood, provided it is allowed enough freedom to capture the  underlying model. A linear model fails to recover the correct cosmology, as it does not have enough freedom to recover the difference between the Ia and non-Ia distribution. It correspondingly has a very high $\chi^2$ relative to the other approaches. The higher-order functions recover consistent cosmologies, and the $\chi^2$ of these models improves by $\Delta \chi^2 < 0.5,$ even though the models have increased the number of parameters by one.

 \begin{table}[htpb!]
\caption{\textbf{Comparison of non-Ia likelihood models for Gaussian simulation:} $\chi^2$ values for the fits using various forms of the non-Ia likelihood for the Gaussian simulations, where the true underlying model is quadratic. The constraints on $\Omega_m,\Omega_\Lambda$ are shown in Figure~\ref{fig:dist}. $\Delta \chi_{\rm eff}^2$ is difference in the effective $\chi^2$ between a given model and the linear case, which has $\chi_{\rm eff}^2 =42526.2 $.\label{table:level1dist}}
\begin{center}
%\vspace{-0.2in}
%\begin{tabular}{p{4cm}p{2.4cm}p{2cm}p{4.9cm}}
\begin{tabular}{c|c|c}
  \hline
  Model & $\Delta \chi_{\rm eff}^2$ & Parameters \\
  \hline
\hline
$  \Upsilon(z)     =  az + c $&0 & 2  \\
\hline
 $   \Upsilon(z)   = az + bz^2 + c$ &-192.9 & 3 \\
\hline
 $    \Upsilon(z)  = az + bz^2 + cz^3 + d$ & -193.3& 4 \\
\hline
$    \Upsilon(z) = {(az+ bz^2 + c)}/{(1+dz)} $  &-193.4 & 4\\
  \hline
  \end{tabular}
\end{center}
\end{table} 

\subsubsection{Dependence on Probability}
The BEAMS algorithm naturally uses some indication of the probability of a data point to belong to the Ia population, whether it is some measure of the goodness-of-fit of the data to a type Ia light-curve template, or something more robust such as the relative probability that the point is a Ia compared to the probability of being of a different type. By including a normalization factor, we can correct for general biases in the probabilities of the Ia points. One might still question, however, how sensitive BEAMS is to the input probability of the objects. 
\begin{figure}[htbp!]
\begin{center}
%\sidecaption[t]
%$\begin{array}{@{\hspace{-0.0in}}l}
\includegraphics[width=0.49\textwidth, trim = 0mm 50mm 0mm 50mm, clip]{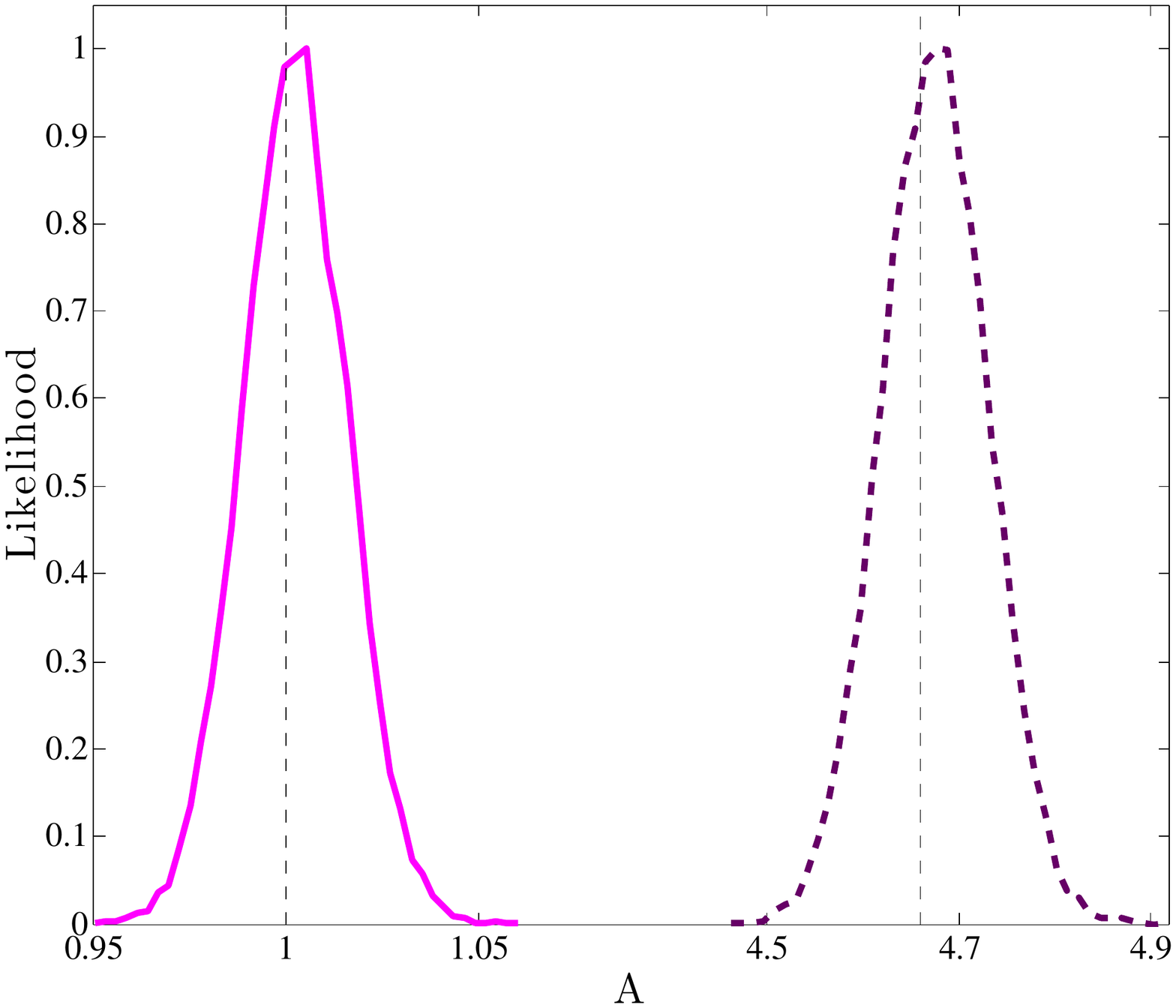}
\includegraphics[width=0.49\textwidth, trim = 0mm 50mm 0mm 50mm, clip]{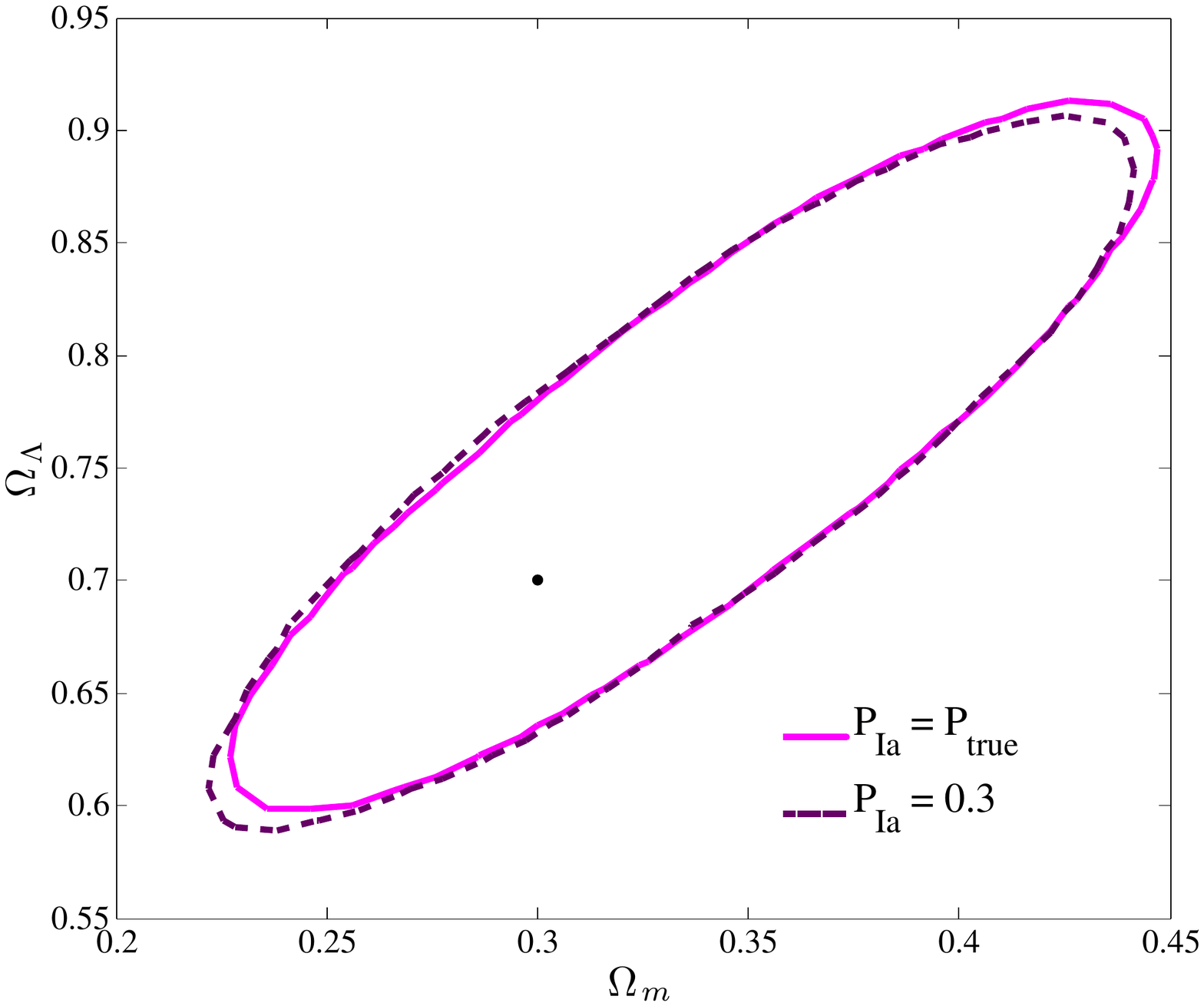}
% \end{array}$
 \caption{\textbf{BEAMS corrects for biased input probability}: the marginalized one-dimensional likelihood for the normalization parameter $A$ (top panel) and estimated contours (bottom panel) for Level I Gaussian simulation under two forms of the probability distribution. The pink curve and contours correspond to the nominal case, where the probabilities are generated in a linear model, and the types are assigned according to the probabilities. The purple dashed contours correspond to assigning a probability of $P_{\mathrm{Ia}} = 0.3$ to all points. The dashed vertical lines show the expected value of the parameter $A$ such that the true input mean probability of $P_{\mathrm{Ia}} = 0.667$ is recovered. Note that the $x$-axis in the top panel has been shortened to allow for comparison of the two distributions. \label{fig:probcheck}}
 \end{center}
 \end{figure}
For the Gaussian simulation, where we assign the probabilities, $P_\mathrm{Ia},$  we can directly change the relationship between the true \textit{underlying} distribution of the types (i.e. the ratio of Ias to non-Ias in the sample) and the \textit{input probability value} (the number we input into the BEAMS algorithm as the $P_\mathrm{Ia}$). If the probabilities are unbiased then the distribution of types should follow the probability distribution of the data, in other words $60\%$ of the points with $P_\mathrm{Ia}=0.6$ should be Type Ia SNe. This is the standard case. We then modify the probabilities by assigning a probability of $P_{\mathrm{Ia}} = 0.3$ to all points (which we know will be biased since the mean probability of the sample is 0.667). 

We compare the constraints in the two cases in Figure~\ref{fig:probcheck}. If we ignore all probability information and set it to a (biased) value of $P_{\mathrm{Ia}} = 0.3,$ the probability information is essentially controlled by the normalization parameter. $A$ tends to a value of 4.7, which, when inserted into Equation~(\ref{eq:aparam}) yields a `normalized' probability of $P_{\mathrm{Ia}} = 0.668$. Hence BEAMS uses the normalization parameter to remap the mean of the \textit{given} probabilities to ones that have a mean that fits the true \textit{unbiased} probabilities. In correcting for this effect, BEAMS manages to recover cosmological parameters consistent with the unbiased case.

\section{Results from the SDSS-II SN data}\label{sec:sdss}

The Sloan Digital Sky Survey Supernova Search operated for three, three-month long seasons during 2005 to 2007. We use the photometric supernovae from all three seasons of the SDSS-II SN survey which also had host galaxy redshifts from the SDSS survey. The analysis and cosmological interpretation of the first season of data (hereafter Fall 2005) are described in \cite{frieman:etal2008, kessler:etal2009,lampeitl:etal2009} and \cite{sollerman:etal2009}. The SDSS CCD camera is located on a 2.5 m telescope at the Apache Point Observatory in New Mexico. The camera operated in the five Sloan optical bands ${ugriz}$ \cite{fukugita:etal1996}. The telescope made repeated drift scans of Stripe 82, a roughly 300 square degree region centered on the celestial equator in the Southern Galactic hemisphere, with a cadence of roughly four to five days, accounting for problems with weather and instrumentation. 

The images were scanned and objects were flagged as candidate supernovae \cite{masao_typer}. Candidate light-curves were compared to a set of supernova light-curve templates in the $g,r,i$ bands (consisting of both core-collapse and Type Ia supernovae) as a function of redshift, intrinsic luminosity and extinction. Likely SNIa candidates were preferentially followed up with spectroscopic observations of both the candidates and their host galaxies (where possible) on various larger telescopes (see \cite{masao_typer}).

In addition to the spectroscopically confirmed SNeIa discovered in the SDSS-II SN, many high-quality candidates without spectroscopic confirmation (i.e. only photometric observations were made of the SNe) but which, by chance, have a host galaxy spectroscopic redshift, are present in the SDSS sample\footnote{The BOSS survey recently obtained host galaxy redshifts of all high-quality SN candidates from all three seasons of the SDSS-II Supernova Search. This work does not use the additional BOSS information and only uses the host galaxy redshifts obtained during the running of the SDSS-II survey. }. 

We include these SNe in both the \textit{cut} sample and the full \textit{photo} sample, but do not set the probabilities of the spectroscopically confirmed \textit{spectro} sample points to unity in the latter. These supernovae are fit with the MLCS2k2 model \cite{jha:etalMLCS2k2} to obtain a distance modulus for each supernova, \textit{assuming} the supernova is a type Ia. 

As outlined in Section~\ref{cutssection}, we impose the standard selection cuts on the probability of the fit to the MLCS2k2 light-curve template ${P}_\mathrm{fit} >  0.01$ and $\Delta >  -0.4$ to all data, and require that the data used have spectroscopic host galaxy redshift information. Applying these cuts to the full three year data yields a photometric sample of 792 SNe, with a spectroscopic subsample of 297 SNe. The \textit{spectro} sample consists of the objects which have been spectroscopically confirmed by other ground-based telescopes, while the \textit{cut} sample consists of the data points which have a typer probability of $P_\mathrm{typer} > 0.9$ and a goodness-of-fit to the light-curve templates within the PSNID typer \cite{masao_typer,sako/etal2011_typer2}, $\chi^2_{lc} < 1.8.$  

\begin{figure}[htbp!]
\begin{center}
\includegraphics[ scale = 0.45, trim = 0mm 50mm 0mm 60mm, clip]{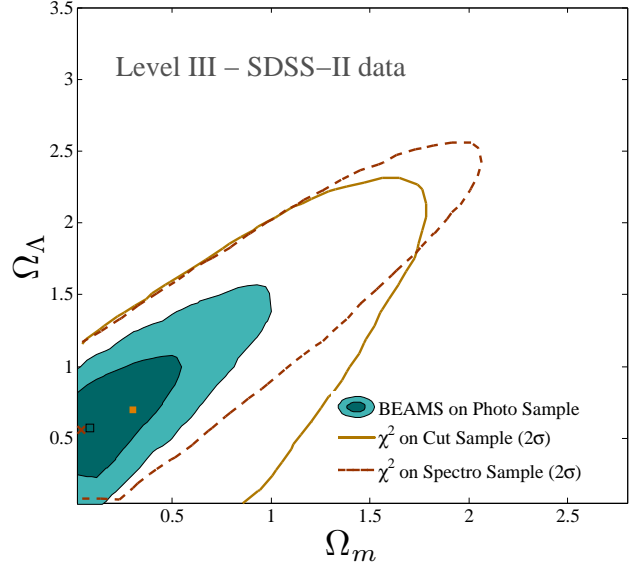}\\[0.0cm]
%,trim = 0mm 40mm 0mm 60mm, clip
 \caption{\textbf{Analysis of the SDSS-II SN data:} We show the $2\sigma$ contours in the $\Omega_m, \Omega_\Lambda$ plane for the SDSS-II SN data. All the constraints --  BEAMS (filled contours with best fit indicated by black open square), the `spectroscopic' sample (dashed contours, best-fit indicated by brown cross) and the `cut' sample -- are consistent with the concordance cosmology (filled brown square). The best-fit BEAMS point is given by the black square, while the best-fit cosmology from the spectroscopic data is indicated by the brown cross. BEAMS provides the smallest contours on the SDSS-II data set, while still being consistent with the constraints from the spectroscopic subsample.
 \label{fig:contour_sdss}}
\end{center}
 \end{figure}

As is shown in Figure~\ref{fig:contour_sdss}, BEAMS estimates parameters consistent with the \textit{spectro} sample as well as the concordance cosmology in the case of the SDSS-II SN data. Moreover, the BEAMS contours are three times smaller than when using the  \textit{spectro} sample alone. In the Gaussian simulations (see Fig.~\ref{fig:contourI}), the BEAMS contours using all the points are $\simeq16\%$ of the size of the \textit{spectro} sample. This highlights the potential of photometric supernova cosmology to drastically reduce the size of error contours with larger samples while remaining unbiased relative to the `known' spectroscopic case.

\section{Conclusions and outlook}

Bayesian Estimation Applied to Multiple Species (BEAMS) is a statistically robust method for parameter estimation in the presence of contamination. The key power of BEAMS is in the fact that it makes use of all available data, hence reducing the statistical error of the measurement, whether or not the purity of the sample can be guaranteed. Rather than discarding data, the probability that the data are ``pure'' is used as a weight in the full Bayesian posterior, reducing potential bias from the interloper distribution.

Here we have presented the algorithm, and discussed in some detail the role of the probabilities. We have tested BEAMS on an ideal Gaussian simulation of two populations and demonstrated that it recovers the input parameters. We have also shown that the BEAMS errors scale as expected with sample size, and that it provides smaller errors than some of the traditional approaches. Using the Gaussian simulation we have further verified that we can detect the correct form of the non-Ia likelihood and correct for a bias in the probabilities.

We have then applied BEAMS to the SDSS-II SN data set of 792 SNe, using photometric data points with host galaxy spectroscopic redshifts, and showed that the BEAMS contours are three times smaller than those obtained when using only the spectroscopically confirmed sample of 297 SNe Ia.

We have restricted ourselves to the binomial case of a SN-Ia population and one general core-collapse, or non-Ia, population. While this assumption is valid for the SDSS-II SN data, we expect that for larger samples a more complicated model with at least two separate non-Ia Gaussians is more appropriate. On the other hand, large supernova surveys will not only increase the total number of type Ia SNe candidates, but will also allow to investigate systematics about the SNe populations directly. The BEAMS algorithm is designed to include and adapt to information about the non-Ia population easily. By adapting the form of the non-Ia population, and including more than one population group, one could use BEAMS to gain insight into the contaminant distribution. 

As we move into the era of huge astronomical surveys that will provide data on thousands of supernovae, BEAMS provides a platform to learn more about the SN populations while at the same time tackling the fundamental questions about the constituents of the universe.

\begin{acknowledgement}
We thank Michelle Knights and the SDSS-II SN team (especially Rick Kessler, John Marriner and Masao Sako) for helpful comments. RH thanks Jo Dunkley, Olaf Davis, David Marsh, Sarah Miller and Joe Zuntz for useful discussions, and thanks the Kavli Institute for Cosmological Physics, Chicago, the South African Astronomical Observatory, the University of Cape Town, and the University of Geneva for hospitality while this work was being completed. MK would like to thank AIMS for hospitality during part of the work. RH acknowledges funding from the Rhodes Trust and Christ Church. MK acknowledges funding by the Swiss NSF. BB acknowledges funding from the NRF and DST. Part of the numerical calculations for this paper were performed on the Andromeda cluster of the University of Geneva.

Funding for the SDSS and SDSS-II has been provided by the Alfred P. Sloan Foundation, the Participating Institutions, the National Science Foundation, the U.S. Department of Energy, the National Aeronautics and Space Administration, the Japanese Monbukagakusho, the Max Planck Society, and the Higher Education Funding Council for England. The SDSS Web Site is http://www.sdss.org/. The SDSS is managed by the Astrophysical Research Consortium for the Participating Institutions. The Participating Institutions are the American Museum of Natural History, Astrophysical Institute Potsdam, University of Basel, University of Cambridge, Case Western Reserve University, University of Chicago, Drexel University, Fermilab, the Institute for Advanced Study, the Japan Participation Group, Johns Hopkins University, the Joint Institute for Nuclear Astrophysics, the Kavli Institute for Particle Astrophysics and Cosmology, the Korean Scientist Group, the Chinese Academy of Sciences (LAMOST), Los Alamos National Laboratory, the Max-Planck-Institute for Astronomy (MPIA), the Max-Planck-Institute for Astrophysics (MPA), New Mexico State University, Ohio State University, University of Pittsburgh, University of Portsmouth, Princeton University, the United States Naval Observatory, and the University of Washington.
\end{acknowledgement}

\bibliographystyle{spphys}
\bibliography{refs_mk}

\end{document}